\documentclass[9pt]{article}

\usepackage{bm}
\usepackage[letterpaper,hmargin=1in,vmargin=1in]{geometry}
\usepackage{times}
\usepackage{multirow}
\usepackage{dcolumn}
\usepackage{tabularx}
\usepackage{graphicx}
\usepackage{subfigure}
\usepackage{ae}
\usepackage{lettrine}
\usepackage{color}
\usepackage{amsmath,amsthm,amssymb,amsfonts,boxedminipage}
\usepackage{epstopdf}
\newcommand\address[1]{\hskip2.25pc \parbox{.8\textwidth}{ \noindent%
   \footnotesize \it \begin{center} #1 \end{center}\rm }  \normalsize \vskip-.2cm }

\renewcommand\title[1]{\bf \hskip2.25pc \parbox{.8\textwidth}{ \noindent%
   \LARGE \bf \begin{center} #1 \end{center} \rm } \vskip.1in \rm\normalsize }

\renewcommand\author[1]{\hskip2.25pc \parbox{.8\textwidth}{ \noindent%
   \normalsize \bf \begin{center} #1 \end{center}\rm } \vskip-1pc }
\makeatletter
\thm@headfont{\sc}
\makeatother

\makeatletter
\renewcommand\@biblabel[1]{#1.}
\renewcommand{\thesubfigure}

\begin{document}

\title{Bayesian machine learning for Boltzmann machine in quantum-enhanced feature spaces}
\author{Yusen Wu$^{1,2,3}$, Chao-hua Yu$^{4}$, $^{\dagger}$Sujuan Qin$^{1}$, \\$^{\ddagger}$Qiaoyan Wen$^{1}$, and $^{*}$Fei Gao$^{1,3}$ }
\address{$^{1}$State Key Laboratory of Networking and Switching Technology,
Beijing University of Posts and Telecommunications, Beijing, 100876, China\\
$^{2}$State Key Laboratory of Cryptology, P.O. Box 5159, Beijing, 100878, China\\
$^{3}$Center for Quantum Computing, Peng Cheng Laboratory, Shenzhen 518055, China\\
$^{4}$The School of Information Technology, Jiangxi University of Finance and Economics, Nanchang 330032, China\\
$\ast$ gaof@bupt.edu.cn, $\dagger$qsujuan@bupt.edu.cn, $\ddagger$wqy@bupt.edu.cn}

\begin{quote}
Bayesian learning is ubiquitous for implementing classification and regression tasks, however, it is accompanied by computationally intractable limitations when the feature spaces become extremely large. Aiming to solve this problem, we develop a quantum bayesian learning framework of the restricted Boltzmann machine in the quantum-enhanced feature spaces. Our framework provides the encoding phase to map the real data and Boltzmann weight onto the quantum feature spaces and the training phase to learn an optimal inference function. Specifically, the training phase provides a physical quantity to measure the posterior distribution in quantum feature spaces, and this measure is utilized to design the quantum maximum a posterior (QMAP) algorithm and the quantum predictive distribution estimator (QPDE). It is shown that both quantum algorithms achieve exponential speed-up over their classical counterparts. Furthermore, it is interesting to note that our framework can figure out the classical bayesian learning tasks, i.e. processing the classical data and outputting corresponding classical labels. And a simulation, which is performed on an open-source software framework for quantum computing, illustrates that our algorithms show almost the same classification performance compared to their classical counterparts. Noting that the proposed quantum algorithms utilize the shallow circuit, our work is expected to be implemented on the noisy intermediate-scale quantum (NISQ) devices, and is one of the promising candidates to achieve quantum supremacy.
\date{\today}
\end{quote}
\quad Quantum computing makes use of the quantum mechanical phenomena, such as quantum superposition and quantum entanglement, to perform computing tasks by running quantum algorithms \cite{Nielsen2002Quantum}. Areas in which quantum algorithms can be applied include simulating quantum systems \cite{Bennett2001Optimal,Childs2012Hamiltonian,Low2017Optimal}, factoring large integer numbers \cite{Shor1997Polynomial} and unstructured database searching \cite{Grover1996A}. And quantum algorithms achieve significant speed-up over their classical counterparts in these areas.

In the past decades, quantum machine learning (QML) became a booming research field attracting worldwide attentions \cite{Wittek2014Quantum}.
It explores how to implement quantum algorithms that could enable machine learning faster than that in classical computers.
Since the proposal of the pioneering quantum algorithm for solving linear systems by Harrow et al.\cite{Harrow2009Quantum}, a variety of quantum algorithms have been proposed to tackle various well-known machine learning problems such as linear regression \cite{Schuld2016Prediction, Yu2017Quantum, Wiebe2012QuantumDataFit, Liu2017QuantumLQR}, data classification \cite{Rebentrost2014Quantum, Wu2019QCRF}, clustering analysis \cite{Lloyd2013QuantumCluster}, principle component analysis \cite{Lloyd2013Quantum} and optimization method \cite{Rebentrost2016Quantum}.
%

Recently, to reduce the impact caused by decoherence, classical-quantum hybrid algorithms are delivered. This kind of algorithm involves minimizing a cost function that depends on the parameters of a quantum gate sequence. Cost evaluation occurs on the quantum computer, with speed-up over classical evaluation, and the classical computer utilizes this cost information to adjust the parameters of the ansatz with the help of suitable classical optimization algorithms. The hybrid algorithms have been proposed for quantum variational eigensolver \cite{MYung2014QVE, Kandala2018QVE}, combinatorial optimization\cite{Farhi2014QAOA, Farhi2016QAOA,Otterbach2017QAOA}, circuit learning \cite{Mitarai2018QCL}, and quantum state diagonalization \cite{Ryan2019VQSD}, in which the hybrid algorithm illustrates quantum advantages.

Some classical-quantum hybrid algorithms reveal significant quantum advantages in feature spaces when handling supervised machine learning tasks. Utilizing the quantum space as the feature space, V. Havlicek et al. \cite{NISQSVM} proposed two quantum algorithms, i.e., a variational quantum classifier and a quantum kernel estimator, which illustrate quantum advantages in terms of the representative power and complexity, respectively. And M. Schuld et al. \cite{Maria2019MLinQspace} used a variational quantum circuit as a linear model to classify data explicitly, which achieves significant speed-up compared with its classical counterparts in the quantum feature space.

It is interesting to note that the above two algorithms \cite{NISQSVM, Maria2019MLinQspace} concentrate on the deterministic task, and these algorithms can be recognized as the quantum counterpart of support vector machine (SVM) algorithm in the quantum feature spaces. Inspired by these two works, we in this paper design a quantum bayesian framework of the restricted Boltzmann machine (RBM), in which the predictions of machine learning tasks are provided probabilistically. The proposed framework contains the encoding phase and the training phase, in detail the encoding phase aims at mapping the real data and Boltzmann weights onto the special designed quantum feature space, and the training phase provides two quantum algorithms to implement the training task. The construction of the quantum bayeisan framework depends on the supportive of the following three components. First, we reform the quantum feature space provided by V. Havlicek et al., encoding both the observed data and corresponding label into feature states, which constitutes a feature space more suitable for bayesian learning. Second, we design a parallel hardware-efficient ansatz to interpret the relationship between visible nodes and hidden nodes in the (RBM). This technique is inspired by the hardware-efficient ansatz which has been widely utilized to simulate the ground state of a molecule in the quantum variational eigensolver (QVE)\cite{MYung2014QVE, Kandala2018QVE}. The parallel hardware-efficient ansatz can manipulate the $\mathcal{O}(n)$ dimensional vector parameter $\mathcal{\bm{\theta}}$ to approximate a $\mathcal{O}(2^n)$ dimensional quantum state. Third, we find a physical quantity to depict the posterior distribution in the quantum feature spaces, and the measure can be interpreted as an overlap between two density matrixes. The first two terms are utilized to complete the encoding phase, and the last one is used to design two quantum algorithms: (1) the quantum maximum a posterior (QMAP) algorithm to find out the appropriate Boltzmann weight; (2) the quantum predictive distribution estimator (QPDE) to efficiently estimate the quantum predictive distribution in the quantum feature space and present the classification results directly. It is shown that both of the proposed quantum algorithms are exponentially faster than the classical counterparts.

Noting that our quantum algorithms process the classical data set $\mathcal{D}=\{\mathbf{x}^{(l)}\}_{l=1}^N$ as well as output the classical label, and we simulate both algorithms on the open-source software framework, HiQ platform \cite{HiQ}. The simulation results show that the QMAP algorithm achieves a success rate of nearly $100\%$, and the QPDE achieves classification with a success rate of $85\%$. Both of the quantum algorithms show almost the same classification performance compared to their classical counterparts. Considering that RBM can be designed to implement many machine learning tasks, including linear classification model, generative model, neural network and so on \cite{Kulchytskyy2016Quantum}, then the proposed quantum bayesian learning framework can be naturally extended to the quantum Gaussian process given by Zhao et al. \cite{ZhikuanZhao2019QBayesian}, and our work is also expected to be utilized in other quantum machine learning models.
\section*{Results}

\textbf{Restricted boltzmann machine via bayesian machine learning.} We first discribe the restricted Boltzmann machine (RBM) model and the methodology to train it by utilizing bayesian framework \cite{HX2014BayesianRBM}. The RBM model is a two-layer, bipartite neural network, which is a "restricted version" of the Boltzmann machine with only inter-connections between hidden layers and visible layers. The visible layer $\mathbf{v}=(v_1...v_{N_v})\in \{0,1\}^{N_v}$ encodes the input data into a $N_v$ dimensional binary string, and the hidden layer $\mathbf{h}=(h_1...h_{N_h})\in \{0,1\}^{N_h}$ is composed of $N_h$ stochastic binary variables. Then the joint probability of $(\mathbf{v,h})$ is expressed as $p(\mathbf{v,h})=\frac{1}{\mathcal{Z}}\exp\left(-E(\mathbf{v,h})\right)$,
where the potential function $E(\mathbf{v,h})=\sum_{i,j}w_{ij}v_ih_j$.
The Boltzmann weights $\mathbf{w}=\left(w_{ij}\right)_{N_v\times N_h}$ compose a symmetry matrix, and the formalized factor $\mathcal{Z}=\sum_{\mathbf{v,h}}\exp\left(-E(\mathbf{v,h})\right)$.

Given the training data $\mathcal{D}=\{\mathbf{x}^{(l)}\}_{l=1}^N$, the Boltzmann weights can be estimated by utilizing the maximum a posterior (MAP) within the bayesian learning framework \cite{Friedman1997Bayesian}:
\begin{align}
\mathbf{w^*}&=\arg \, \max_{\mathbf{w}}p(\mathbf{w}|\mathcal{D})=\arg \, \max_{\mathbf{w}}p(\mathbf{w})p(\mathcal{D}|\mathbf{w}).
\end{align}
The probability $p(\mathbf{w})$ is the priori distribution of the Boltzmann weights, and $p(\mathcal{D}|\mathbf{w})$ is the likely-hood probability distribution, which can be computed as $p(\mathcal{D}|\mathbf{w})=\prod_{l=1}^N\sum_{\mathbf{h}}p(\mathbf{x}^{(l)},\mathbf{h}|\mathbf{w})$ on the RBM model.

Meanwhile, in some application scenarios, the predictive distribution of the new data point $\mathbf{x}^*$ prevails over the Boltzmann parameter $\mathbf{w}$ itself, and we can directly express the predictive distribution $p(t|\mathbf{x}^{*},\mathcal{D})$ by sum of $p(t|\mathbf{x}^*,\mathbf{w})$ with the weight parameter $p(\mathbf{w}|\mathcal{D})$:
\begin{align}
p(t|\mathbf{x}^{*},\mathcal{D})=\int_{\mathbf{w}}p(t|\mathbf{x^*},\mathbf{w})p(\mathbf{w}|\mathcal{D}){\rm d}\mathbf{w}.
\end{align}
In general, the distribution $p(t|\mathbf{x}^{*},\mathcal{D})$ can not be analytically computed, therefore researchers often utilize Markov Chain Monte Carlo (MCMC) method to estimate this distribution. Nevertheless, if the machine learning tasks process the data on a high dimensional feature space $\Phi(\cdot)$, estimating the posterior distribution $p\left(\mathbf{w}|\Phi(\mathcal{D})\right)$ and predictive distribution $p\left(t|\mathbf{x}^*,\Phi(\mathcal{D})\right)$ will take a large amount of computational overheads which is extremely hard for classical computing. We therefore choose the quantum state space as the feature space, and we design a quantum bayesian framework to handle this issue. \\
\textbf{Encoding phase of bayesian learning framework.} The quantum bayesian framework contains two fundamental components, namely the encoding phase and the training phase. The first stage of encoding phase indicates the procedure that maps the real data onto the quantum feature space which is special designed for quantum bayesian learning.
\begin{figure*}[htp]
  \begin{center}
  \includegraphics[width=1.0\textwidth]{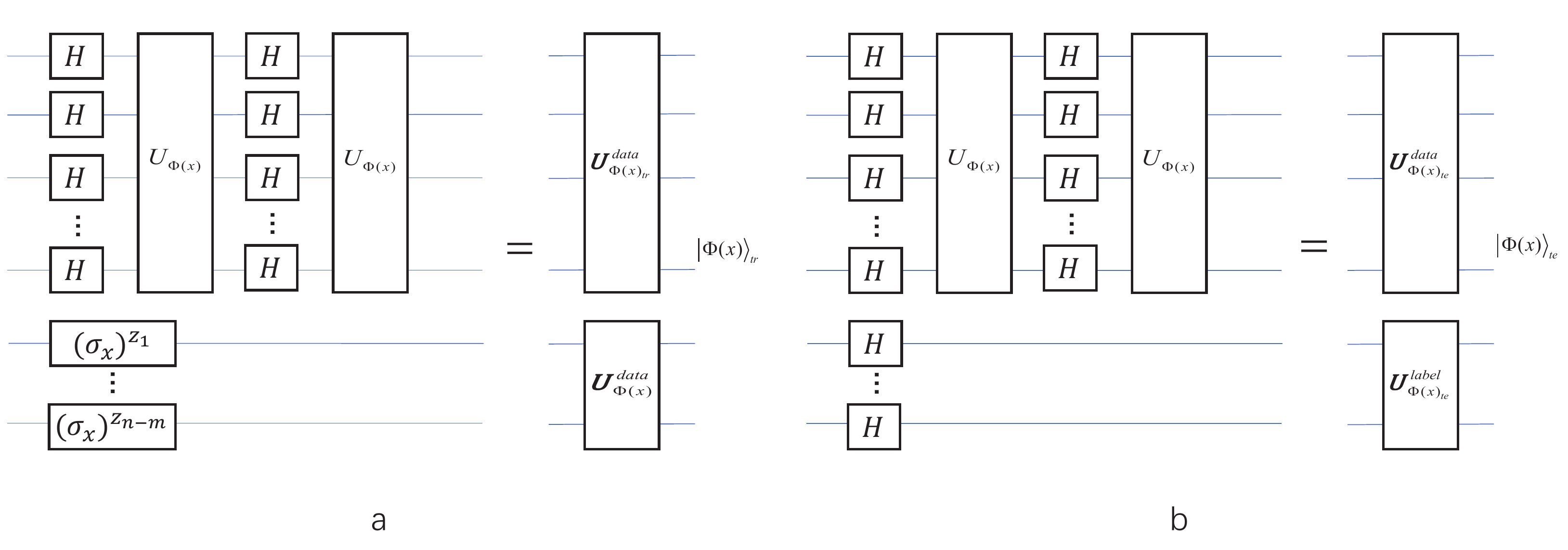}
  \caption{Circuits of generating the feature states $|\Phi(\mathbf{x})\rangle_{tr}$ and $|\Phi(\mathbf{x})\rangle_{te}$, respectively.
  }
  \label{fig:quantum feature space}
  \end{center}
\end{figure*}
In quantum settings, the feature map is an injective encoding of classical information $\mathbf{x}\in \mathcal{R}^d$ into a quantum state $|\Phi\rangle\langle\Phi|$ on an $m-$qubits register, s.t. $\Phi:\mathbf{x}\rightarrow|\Phi(\mathbf{x})\rangle\langle\Phi(\mathbf{x})|$.
The family of feature map circuit can be defined as
$\mathcal{U}_{\mathbf{x}_{data}}=U_{\Phi(\mathbf{x}_{data})}H^{\otimes m}U_{\Phi(\mathbf{x}_{data})}H^{\otimes m}$,
where the unitary operator
\begin{align}
U_{\Phi(\mathbf{x})}=\exp\left(i\sum\limits_{S\subset [m]}\phi_S(\mathbf{x})\prod\limits_{i\in S}\sigma_i^z\right).
\end{align}
The notation $S$ indicates a subset that belongs to set $[m]=\{1,2,...,2^m\}$, and $\phi_S(\mathbf{x})\in \mathcal{R}$ are non-linear functions of the input data $\mathbf{x}\in \mathcal{R}^d$ \cite{NISQSVM}, and other choices of the feature maps are also possible such as squeezed vacuum state \cite{Maria2019MLinQspace}. For the supervised classification tasks, the training data $\mathbf{x}$ is stored into two parts $\mathbf{x}=(\mathbf{x}_{data},\mathbf{x}_{label})$, namely data parts and label parts. Then the feature state is defined as:
\begin{align}
|\Phi(\mathbf{x})\rangle_{tr}=\mathcal{U}_{\mathbf{x}_{data}}|0\rangle^{\otimes m}\mathcal{U}_{\mathbf{x}_{label}}|0\rangle^{\otimes n-m},
\end{align}
where the operator $\mathcal{U}_{\mathbf{x}_{label}}=(\sigma_x)^{z_1}...(\sigma_x)^{z_{n-m}}$ encodes the label parts into the last $n-m$ qubits, and the binary string $(z_1...z_{n-m})$ indicates the $\mathbf{x}_{label}$.  For instance, if handling the binary classification task, the label is marked by $\mathcal{U}_{\mathbf{x}_{label}}|0\rangle=\sigma_x|0\rangle=|1\rangle$ when $\mathbf{x}_{label}=1$, otherwise $\mathcal{U}_{\mathbf{x}_{label}}|0\rangle=|0\rangle$.

On the other hand, the predictive data $\mathbf{x}$, which does not have its label, is encoded by the quantum feature map:
\begin{align}
|\Phi(\mathbf{x})\rangle_{te}=\mathcal{U}_{\mathbf{x}_{data}}|0\rangle^{\otimes m}H^{\otimes n-m}|0\rangle^{\otimes n-m}.
\end{align}
Still taking the binary classification into consideration, the feature state of predictive data $\mathbf{x}$ becomes $\mathcal{U}_{\mathbf{x}_{data}}|0\rangle^{\otimes m}|+\rangle$. Fig.\ref{fig:quantum feature space} illustrates the quantum circuits of these two quantum feature maps respectively.

The second subroutine aims at encoding the parameter $\mathbf{w}$ into a density matrix $\rho(\mathbf{w})$, whose structure satisfies connection rules between the visible nodes and hidden nodes. To achieve this, we propose a parallel hardware-efficient ansatz to represent the priori density matrix $\rho(\mathbf{w})$ of the RBM model (see Methods).\\
\noindent \textbf{Training phase of quantum bayesian framework.} We then discuss the training phase of our framework. First, we deliver a quantum algorithm for maximum the posterior distribution, and this quantum algorithm is designed according to the measure of a physical quantity. We find the overlap between the priori density matrix $\rho(\mathbf{w})$ and likelyhood density matrix $\rho(\Phi(\mathcal{D})|\mathbf{w}(\bm{\theta}))$ can be recognized as a principle to evaluate the posterior distribution in quantum settings, furthermore this physical quantity can be efficiently computed by quantum computer but $\#P$ hard for classical computer \cite{G2017Complexity}, and this measure reveals the quantum advantages.
The validity for choosing this measure as well as the method for computing the likelyhood density matrix $\rho(\Phi(\mathcal{D})|\mathbf{w}(\bm{\theta}))$ are presented in Methods.

Specifically, the quantum maximum posterior distribution (QMAP) algorithm can be regarded as a procedure that adjusts the parameter $\bm{\theta}$ to achieve
\begin{align}
\label{eq:QMAP}
\bm{\theta^*}=\arg\max_{\bm{\theta}}{\rm Tr}\left(\rho(\Phi(\mathcal{D})|\mathbf{w}(\bm{\theta}))\rho(\mathbf{w}(\bm{\theta}))\right).
\end{align}
The QMAP algorithm starts with estimating the measure, s.t. ${\rm Tr}\left(\rho(\Phi(\mathcal{D})|\mathbf{w}(\bm{\theta}))\rho(\mathbf{w}(\bm{\theta}))\right)$, by swap test technique \cite{HBuhrman2001Fingerprint}. Utilizing this information, QMAP algorithm is followed by simultaneous perturbation stochastic approximation (SPSA) approach \cite{Spall1992SPSA} to find out the optimal parameter $\bm{\theta}$. For every step $k$ of the optimization, we sample from $p$ symmetrical Bernoulli distributions $\Delta_k$ using preassigned elements from a sequence converging to zero, s.t., $c_k$. The gradient at $\bm{\theta_k}$ is approximated using energy evaluations at $\bm{\theta}_k^{\pm}=\bm{\theta}_k\pm c_k\Delta_k$, and is constructed as
\begin{align}
\label{eq:SPSA}
g_k(\bm{\theta}_k)=\frac{1}{2c_k}\left({\rm Tr}\left(\rho(\Phi(\mathcal{D})|\mathbf{w}(\bm{\theta}_k^+))\rho(\mathbf{w}(\bm{\theta}_k^+))\right)-
{\rm Tr}\left(\rho(\Phi(\mathcal{D})|\mathbf{w}(\bm{\theta}_k^-))\rho(\mathbf{w}(\bm{\theta}_k^-))\right)\right)\Delta_k.
\end{align}
Noting that this gradient approximation only requires two estimations of the energy, regardless of the number of variables in $\bm{\theta}$. The value of $g_k(\bm{\theta}_k)$ can be obtained by swap test technique, then the parameters are updated as
\begin{align}
\bm{\theta}_{k+1}=\bm{\theta}_k+\eta g_k(\bm{\theta}_k).
\end{align}
The step length $\eta$ is selected based on experience, and the optimal solution $\bm{\theta}^*$ convergences rapidly after several iterations in this iterative algorithm. So far, the QMAP algorithm obtains the optimal parameter $\bm{\theta}^*$ which can be used to classify the predictive data.

The classification rules of QMAP algorithm are interpreted as follows. After determining the optimal solution $\bm{\theta}^*$, the likelyhood distribution $\rho(t|\mathbf{x}^*,\mathbf{w})$ of a test data point $\mathbf{x}^*$ can be computed by the QCL subroutine (see Methods), where the training feature state $|\Phi(\mathbf{x}^*)\rangle_{tr}$ should be substituted by the testing feature state $|\Phi(\mathbf{x}^*)\rangle_{te}$. Then we utilize computational basis to measure the last $n-m$ registers of $\rho(t|\mathbf{x}^*,\mathbf{w})$. Taking binary classification as an example, we select $\{|0\rangle,|1\rangle\}$ basis to measure the third register. If the probability $p(0)>p(1)$, then $\mathbf{x}^*$ is assigned to the class $-1$, otherwise $+1$.

The QMAP algorithm involves swap test technique as well as the QCL subroutine proposed in Methods, which actually takes $\mathcal{O}\left(\frac{MN\rm{poly}(n)}{\varepsilon^2}\right)$ gate complexity. The parameter $n$ is the number of utilized qubits or the scale of the quantum system, $\varepsilon$ indicates the error taken by swap test, $N$ represents the number of data points in the data set, and $M$ is the iterative times of SPSA approach. Given the real data set $\mathcal{D}=\{\mathbf{x}^{(l)}\}_{l=1}^N, \mathbf{x}^{(l)}\in \mathcal{R}^d$ and the corresponding feature space $\Phi(\mathcal{D})=\{\Phi(\mathbf{x}^{(l)})\}_{l=1}^{N}, \Phi(\mathbf{x}^{(l)})\in \mathcal{R}^{2^d}$, then the QMAP algorithm achieves the exponential speed-up under the assumption of $d=\mathcal{O}(n)$.\\
\begin{figure}[htp]
  \begin{center}
  \includegraphics[width=1\textwidth]{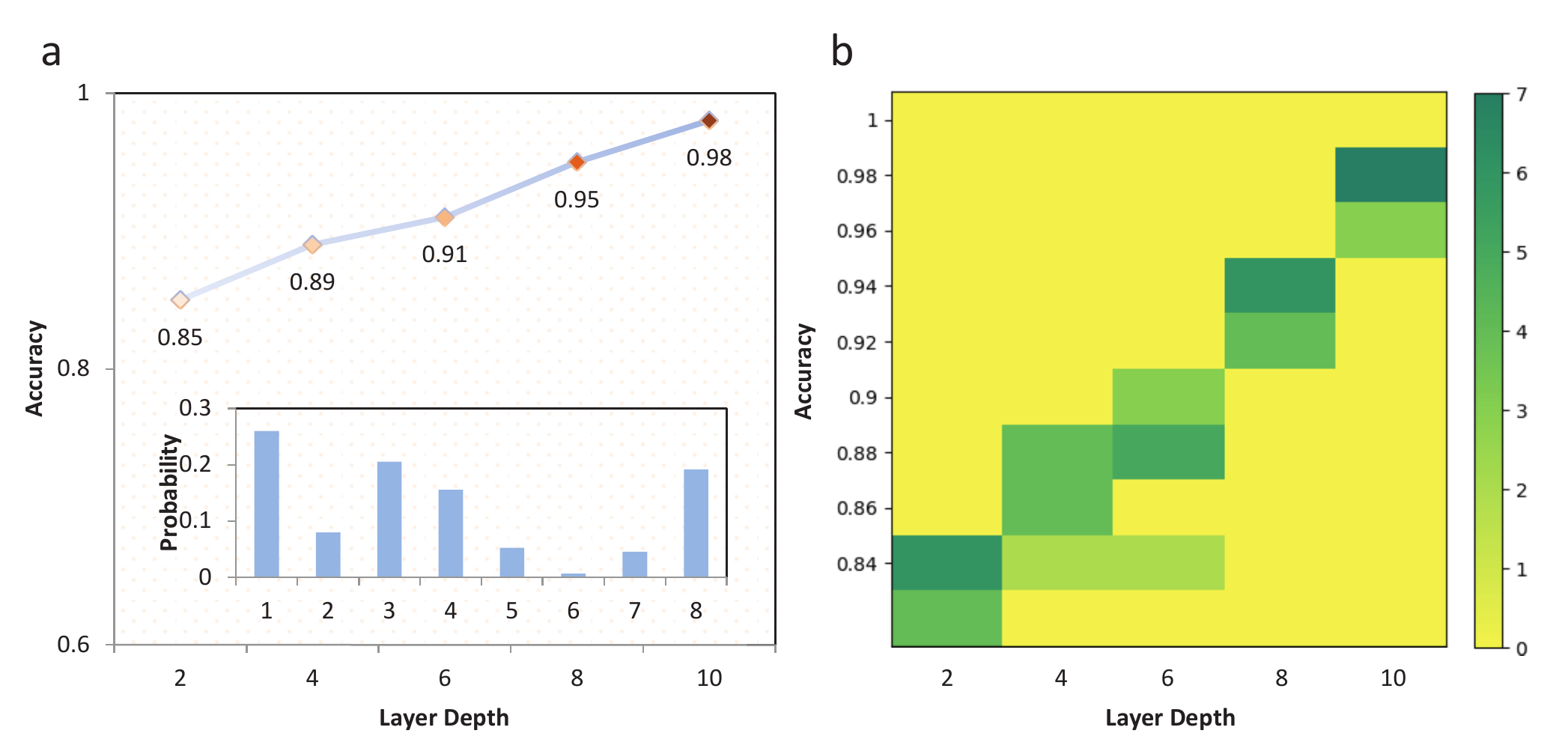}
  \caption{(a) Experimental classification result by QMAP algorithm. The blue line chart shows the classification accuracies for different depths, and each depth involves training 400 data points and testing 40 data points for 10 times, where the mean values for the 10 times are represented by the square-shaped filled dots. And the inset blue bars shows the probability distribution of $|\mathbf{w}\rangle$ with depth 10 and accuracy $98.1\%$. (b) The accuracy distribution of 10 experimental results for each depth.}
  \label{fig:experimentla results of QMAP}
  \end{center}
\end{figure}
\begin{figure*}[htp]
  \begin{center}
  \includegraphics[width=1\textwidth]{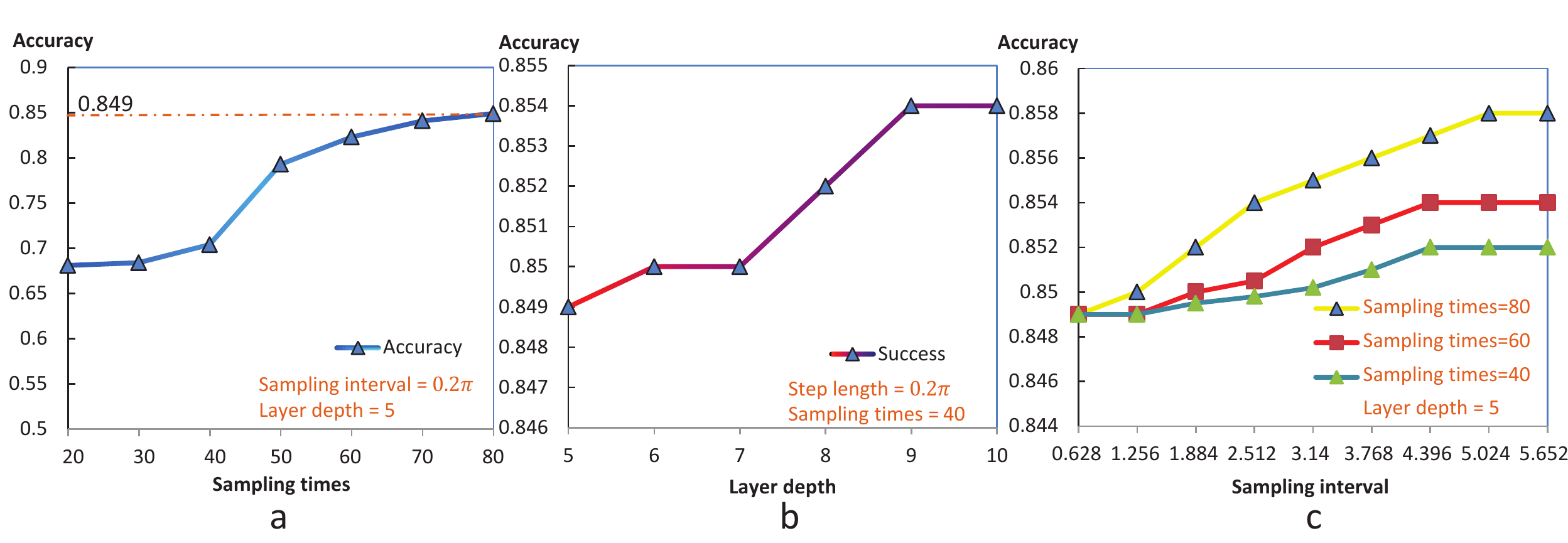}
  \caption{(a) Illustration of the classification accuracy utilizing QPDE, the number of sampling times vary from 20 to 80 with variation length 20. The sampling interval of parameter $\theta$ is $0.2\pi$, and the depth of variational quantum circuit for generating state $|\mathbf{w}\rangle$ equals to 5. (b) Illustration of the classification success with the changing of layer depth, from 5 to 10, the success curve witnesses from 0.849 to 0.854. (c) Success rates by QPDE for sampling 40, 60, 80 times, respectively.}
  \label{fig:experimental results of QPDE}
  \end{center}
\end{figure*}

Then we discuss another inference method in our framework. The quantum predictive distribution estimator (QPDE) can directly construct a predictive distribution rather than deliver the Boltamann weight, and this result naturally corresponds to a quantum state which represents such distribution. Given a predictive visible data $\mathbf{x}^*$ (without label), the QPDE aims at obtaining the corresponding predictive distribution
\begin{align}
\label{eq:predictive-distribution}
\rho(t|\mathbf{x}^{*},\Phi(\mathcal{D}))=\int_{\mathbf{w}}\rho(t|\mathbf{x^*},\mathbf{w}){\rm Tr}\left(\rho(\Phi(\mathcal{D})|\mathbf{w})\rho(\mathbf{w})\right){\rm d}\mathbf{w},
\end{align}
where $\rho(t|\mathbf{x^*},\mathbf{w})$ is the likelyhood distribution of the predicted data $\mathbf{x}^*$ that has been introduced above. We indicate the fact that traversing all the possible state $|\mathbf{w}\rangle$ can be achieved by modifying the rotation parameter $\bm{\theta}$ accompanied by an appropriate error in Theorem 2 (see Methods). Generally, the predictive distribution $p(t|\mathbf{x}^{*}, \mathcal{D})$ can not be calculated analytically, unless priori and likelyhood distribution belong to the same conjugate distribution family. Furthermore, the likelyhood density matrix $\rho(t|\mathbf{x}^{*},\Phi(\mathcal{D}))$, in which visible data has been mapped on the quantum feature space, is computationally hard by classical means because of the construction of the feature map $|\Phi(\mathbf{x})\rangle$ and overleap $K(\mathbf{x}^{(l)},\mathbf{w})$. Thus we also believe that function (\ref{eq:predictive-distribution}) may not be estimated directly by Markov chain Monte Carlo (MCMC) method efficiently \cite{NISQSVM}.
\begin{figure*}[htp]
  \begin{center}
  \includegraphics[width=0.9\textwidth]{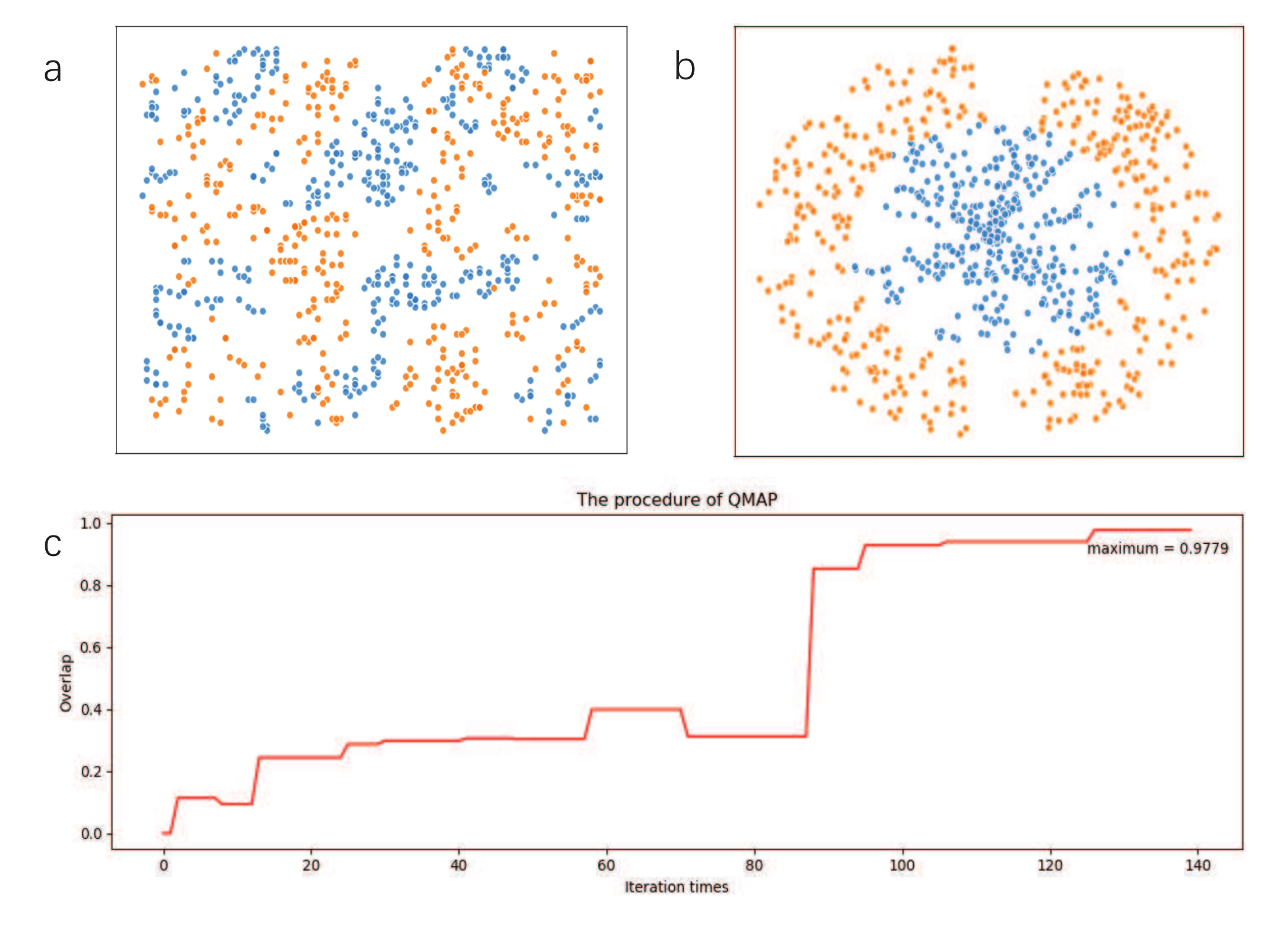}
  \caption{(a) Example data used for QMAP algorithm. The data labels (red for +1 label and blue for -1 label). (b) Decision bound provided by QMAP algorithm with 10 layers depth. (c) Evolution of the quantum posterior distribution ${\rm Tr}\left(\rho(\Phi(\mathcal{D})|\mathbf{w}(\bm{\theta}))\rho(\mathbf{w}(\bm{\theta}))\right)$ after $140$ iterations of using SPSA algorithm. This graph corresponds to the quantum circuit with $10$ layers depth.}
  \label{fig:experimental results of QMAP_2}
  \end{center}
\end{figure*}
Combining MCMC method and quantum technique, the QPDE provides a method to estimate function (\ref{eq:predictive-distribution}). We first select and prepare a sampling distribution $h(\bm{\theta})$ which satisfies the condition $\int_{\bm{\theta}}h(\bm{\theta})d\bm{\theta}=1$, and there are some possible choices of $h(\bm{\theta})$ such as uniform distribution, multivariate Gaussian distribution and multivariate Laplasian distribution. Then the function (\ref{eq:predictive-distribution}) can be rewritten as
\begin{align}
\rho(t|\mathbf{x}^{*},\mathcal{D})\propto\int_{\bm{\theta}}\frac{\rho(t|\mathbf{x^*},\bm{\theta}){\rm Tr}\left(\rho(\Phi(\mathcal{D})|\bm{\theta})\rho(\mathbf{w}(\bm{\theta}))\right){\rm }}{h(\bm{\theta})/n\bm{\theta}^{n-1}}h(\bm{\theta})d\bm{\theta}.
\end{align}
This is followed by obtaining a set of samples $\mathbf{z}^{(l)}$ (where $l=1,2,...,K$) drawn independently from the distribution $h(\bm{\theta})$, which allows the function (\ref{eq:predictive-distribution}) to be approximated by a finite sum
\begin{align}
\label{eq:QMCMC}
\rho(t|\mathbf{x}^{*},\mathcal{D})\propto\frac{1}{K}\sum\limits_{l=1}^Kf(t|\mathbf{z}^{(l)}, \mathcal{D}),
\end{align}
where
\begin{align}
f(t|\mathbf{z}^{(l)}, \mathcal{D})=\frac{\rho(t|\mathbf{x^*},\mathbf{z}^{(l)}){\rm Tr}\left(\rho(\Phi(\mathcal{D})|\mathbf{z}^{(l)})\rho(\mathbf{w}(\mathbf{z}^{(l)})\right){\rm }}{h(\mathbf{z}^{(l)})/n(\mathbf{z}^{(l)})^{n-1}}.
\end{align}
Noting that QPDE does not directly achieve $\rho(t|\mathbf{x}^*,\mathcal{D})$, but it just measures the last $n-m$ qubits and then performs simple calculations on the measurement results. We still deliver a binary classification as an illustrative example. In detail, parameters ${\rm Tr}\left(\rho(\Phi(\mathcal{D})|\mathbf{z}^{(l)})\rho(\mathbf{w}(\mathbf{z}^{(l)})\right)$ can be calculated by the swap test, then one can measure the last qubit of $\rho(t|\mathbf{x^*},\mathbf{w}^{(k)})$ by computational basis $\{|0\rangle,|1\rangle\}$, where the probability $p(0), p(1)$ can be estimated by simply counting the measurement results. Finally $\mathbf{x}^*$ is assigned to the class $-1$ if $p(0)>p(1)$, otherwise $+1$ class is assigned.

Noting that QPDE also needs the supportive of QCL subroutine, meanwhile sampling from the distribution $h(\bm{\theta})$ takes additive complexity of $\mathcal{O}(K)$ times. Therefore, QPDE takes $\mathcal{O}\left(\frac{KN\rm{poly}(n)}{\varepsilon^2}\right)$ time, where $N$ is the number of data points in real dataset, $\varepsilon$ represents the error taken by swap test.


\begin{figure}[htp]
  \begin{center}
  \includegraphics[width=0.8\textwidth]{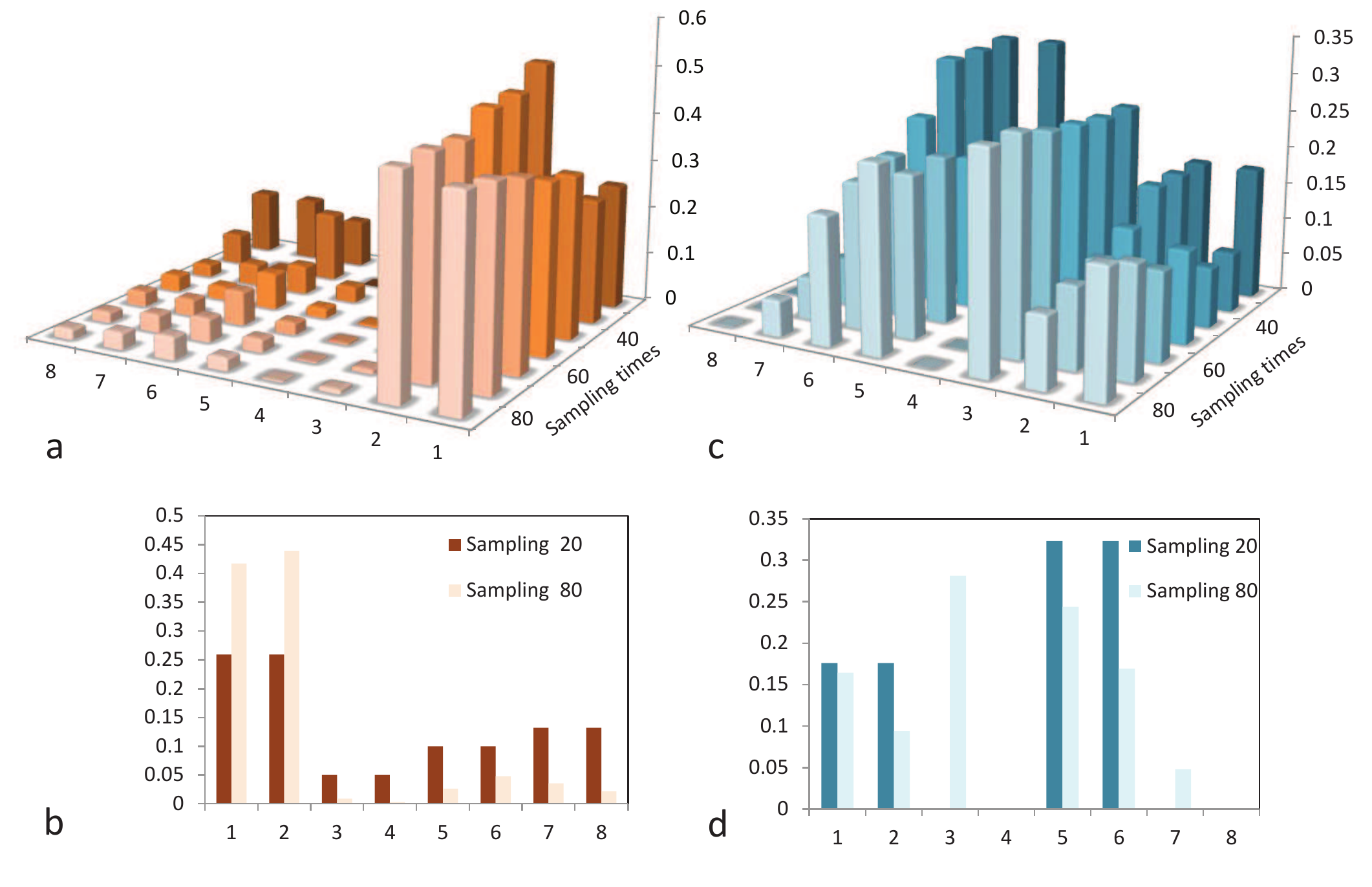}
  \caption{Wave-function evolutions by running QPDE for different sampling times on the two testing data points $\mathbf{x^*}=(4.272566, 5.08938)$ and $\mathbf{y^*}=(5.08938, 3.39293)$, shown in the left side (a and b) and right side (c and d), respectively.
  There are eight amplitudes in the wavefunction and the $\mathbf{x}^*_{label}=+1$ and $\mathbf{y}^*_{label}=-1$.}
  \label{fig:wavefunction}
  \end{center}
\end{figure}
\begin{figure*}[htp]
  \begin{center}
  \includegraphics[width=0.80\textwidth]{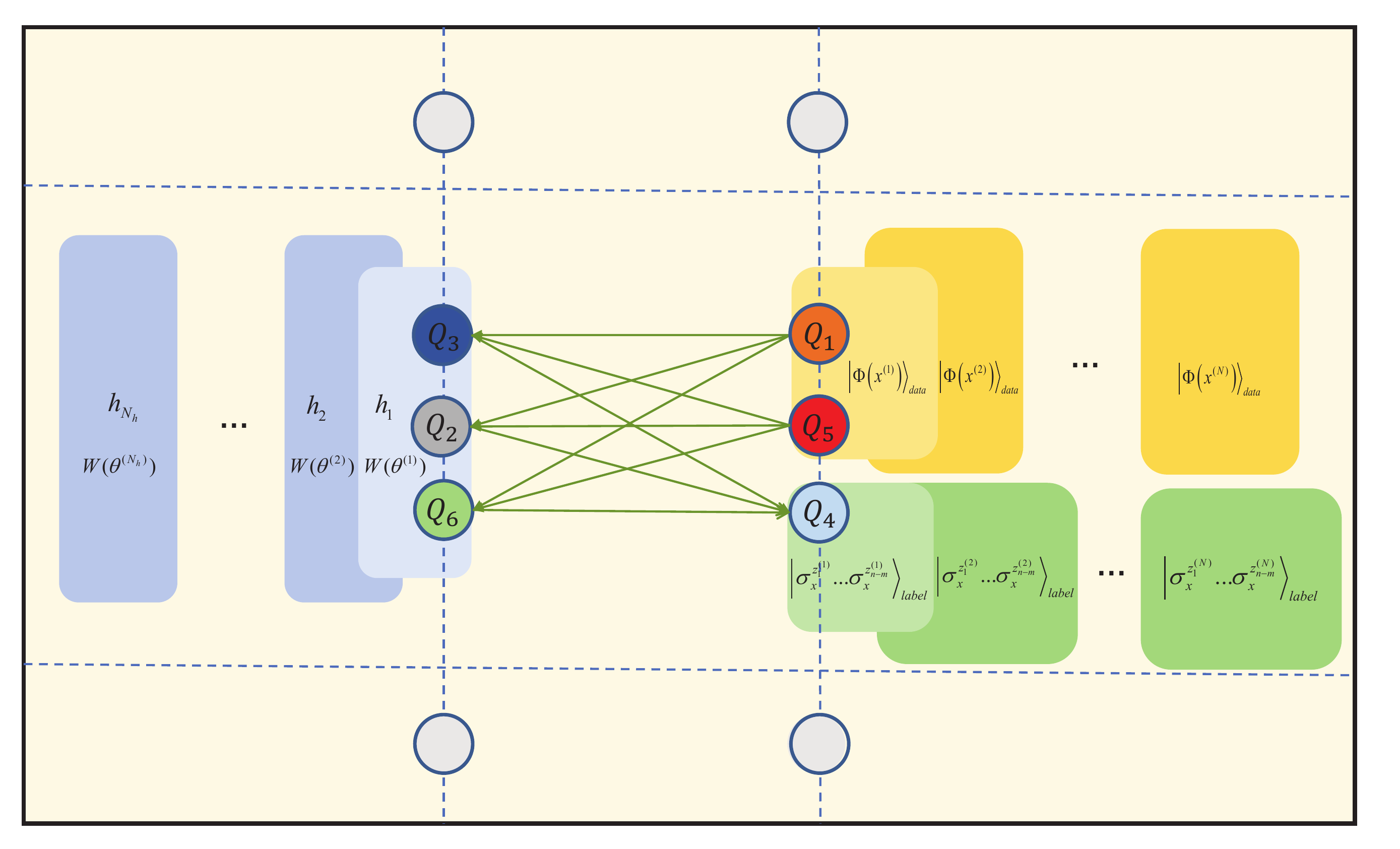}
  \caption{Illustration of the quantum restricted boltzmann machine (QRBM) which is composed of 6 qubits. In detail, qubits $Q_2, Q_3, Q_6$ are utilized to encode hidden nodes $(h_1,...,h_{H_h})$ in the form of superposition $1/\sqrt{N_h}\sum_{i=1}^{N_h}W(\theta^{(i)})|0\rangle_{Q_2Q_3Q_6}$. The qubits $Q_1, Q_5$ and $Q_4$ are the visible nodes, from which $Q_1, Q_5$ encode the visible data via nonlinear operator $\mathcal{U}_{\mathbf{x}_{data}}$ and $Q_4$ represents the corresponding label of $\mathbf{x}$.
  }
  \label{fig:QBM}
  \end{center}
\end{figure*}
\noindent\textbf{Experimental results.} The experimental preparation of the QMAP algorithm is illustrated as follows. We utilize $1$ hidden node $h_1$ and $3$ visible nodes $v_1, v_2, v_3$ to construct a QRBM to implement a binary classification task. At first, we encode the classical training data $\mathcal{D}=\{\mathbf{x}^{(l)}\}_{l=1}^N$ onto the quantum qubits $Q_1, Q_5$ and $Q_4$ by utilizing the quantum feature map $|\Phi(\mathbf{x})\rangle_{tr}=\mathcal{U}_{\mathbf{x}_{data}}|0\rangle_{ Q_1}|0\rangle_{Q_5}\mathcal{U}_{\mathbf{x}_{label}}|0\rangle_{Q_4}$, which is illustrated in Fig.\ref{fig:QBM}. Then we invoke the QCL subroutine to add the hidden node $h_1$ onto the RBM model and computing the likelyhood distribution at the same time. Finally, QMAP algorithm can find out the optimal parameter $\bm{\theta}^*$ with the help of SPSA algorithm and swap test technique, and the evolution of quantum posterior distribution is illustrated as Fig.\ref{fig:experimental results of QMAP_2}(c). The optimal parameter $\bm{\theta}^*$ can be further examined by the testing data set. We set the iterative parameter $c_k=1/k^{0.6}$, afterwards the test results are illustrated as Fig.\ref{fig:experimentla results of QMAP}. Our quantum algorithm yields $98.1\%$ success for the test data set. The original data distribution and the decision boundary of the QMAP algorithm are illustrated as Fig.\ref{fig:experimental results of QMAP_2}(a) and (b).

For the experiments of QPDE, the preparation stage is to obtain a set of samples $\bm{\theta}^{(k)}$ from a uniform distribution $h(\bm{\theta})$ and the function (\ref{eq:predictive-distribution}) can be evaluated as
\begin{align}
\label{eq:QMCMCUni}
\rho(t|\mathbf{x}^{*},D)\approx\sum\limits_{k=0}^{K}\bigtriangleup_{\mathbf{w}}\rho(t|\mathbf{x^*},\mathbf{w}_{(k)}){\rm Tr}\left(\rho(\mathcal{D}|\mathbf{w}_{(k)})\rho(\mathbf{w}_{(k)})\right),
\end{align}
where the parameter $K=\mathcal{O}(1/\varepsilon^{'})$, $\triangle_{\mathbf{w}}=\mathcal{O}(\varepsilon^{'})$, and $|\mathbf{w}_{(k)}\rangle=W\left(\bm{\theta}^{(k)}\right)|0\rangle^{\otimes n}$.
We first implement the quantum variational classifier $\rho(t|\mathbf{x^*},\mathcal{D})$ with $20, 40, 60, 80$ sampling
times, respectively, on the quantum simulation processor. We expect the classification success rate goes up with increasing the number of samplings. We utilize training sets consisting of $400$ data points per label. And we take 40 testing data points for each class. Two of the testing data $\mathbf{x^*}=(4.272566, 5.08938)$ and  $\mathbf{y^*}=(5.08938, 3.39293)$'s wavefunction are shown in Fig.\ref{fig:wavefunction} for different number of samplings. And the classification success is also illustrated in Fig.\ref{fig:experimental results of QPDE}(a). We observe that the upper bound of the accuracy converges to $0.850$ albeit with more optimization steps. The number of layer depth of priori distribution $\rho(\mathbf{w})$ is set to $5$, and sampling interval is $0.2\pi$.

Second, we implement the quantum variational circuit $W(\bm{\theta})$ for different layer depth from $5$ to $10$ on the quantum simulation processor. We observe that the classification accuracy shows a slight rise from $0.850$ to $0.854$ with layer depth varying from $5$ to $10$, and maintains a constant regardless of increasing the layer depth. The sampling interval is set to $0.2\pi$ and the number of sampling times is $40$.

Finally, we also implement the experiments on varying the sampling interval from $0.2\pi$ to $2\pi$ when executing $40$ to $80$ different sampling times on the $5$ layer depth quantum circuit. It is interesting to note that larger number of sampling times with larger $\bm{\theta}$ interval actually yield higher success rate up to $85.8\%$.

According to the experiments results, compared with QPDE, the QMAP algorithm has a larger complexity, but achieves higher success rate up to $98.1\%$. Therefore, we in practice have to face a tradeoff between complexity and success rate before choosing a more suitable one between the two algorithms.
\section*{Discussion}
In summary, we implement a quantum bayesian learning framework in the exponentially large feature spaces. Our framework provides two variational quantum algorithms, which build upon the realization that the posterior and predictive distributions are hard to estimate classically in the quantum-state feature spaces. Both of them can handle the classical dataset and output the classical predictive label, and it is shown that our algorithms are able to be exponentially faster than the classical counterparts. We hope our work can inspire more bayesian machine learning algorithms accessible to NISQ devices.
.

\section*{Methods}
\begin{figure*}[htp]
  \begin{center}
  \includegraphics[width=0.9\textwidth]{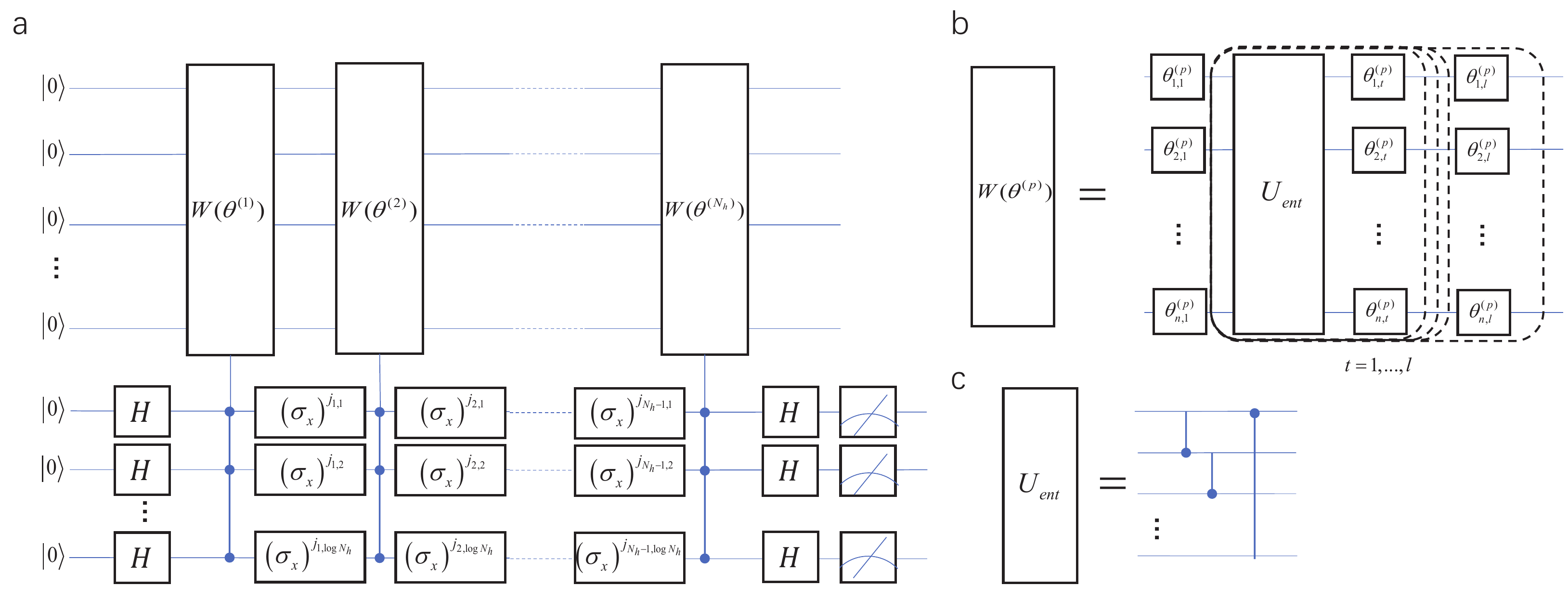}
  \caption{(a) Illustration of the quantum circuit of $W=\sum_{p=1}^{N_h}|p\rangle\langle p|\otimes W(\bm{\theta}^{(p)})$ by introducing $\log N_h$ ancilla qubits. And the gate complexity of $W$ is $\mathcal{O}(N_h\log N_h)$. (b) Variational circuit used for encoding the boltzmann weight between the hidden nodes $N_p$ and visible nodes. We choose the hardware-efficient ansatz for the variational unitary $W(\bm{\theta}^{(p)})=U_{loc}^{(l)}(\theta_l^{(p)})U_{ent}...U_{loc}^{(2)}(\theta_2^{(p)})U_{ent}U_{loc}^{(1)}(\theta_1^{(p)})$, with entanglement layer $U_{ent}$. (c) The fundamental structure of $U_{ent}$.
  }
  \label{fig:parallel hardware efficient ansatz}
  \end{center}
\end{figure*}

\textbf{Parallel hardware-efficient ansatz.} In this section, we present a quantum technique to generate the priori probability in the pattern of quantum state. To process the data in the high dimensional quantum-enhanced feature space with $\mathcal{O}(2^n)$ dimension, researchers propose the hardware efficient ansatz to approximate an arbitrary quantum state with an affordable error \cite{Kandala2018QVE}. To represent the relationship between the visible nodes and hidden nodes, we propose the parallel hardware-efficient ansatz whose fundamental quantum circuit is illustrated in Fig.\ref{fig:parallel hardware efficient ansatz}(a). If the RBM model has $N_h$ hidden nodes $h_1,...,h_{N_h}$, the quantum priori quantum state can be prepared by
\begin{align}
\rho(\mathbf{w})=W(\bm{\theta})\left(|0\rangle\langle0|\right)^{\otimes n}W^{\dag}(\bm{\theta})=|\mathbf{w}\rangle\langle\mathbf{w}|,
\end{align}
in which $W(\bm{\theta})=\sum_{p=1}^{N_h}W(\bm{\theta}^{(p)})$. The shallow quantum circuit $W(\bm{\theta}^{(p)})$ can be achieved by appending layers of single-qubit unitaries and entangling gates, and each layer contains an additional set of entanglers across all the qubits used. The operator $W(\bm{\theta}^{(p)})$ is a circuit of $l$ repeated entanglers, and interleave them with layers comprised of local single qubit rotations:
$U_{loc}^{(t)}(\theta_t^{(p)})=\otimes_{m=1}^{n}\left(e^{i\left(\theta_{m,t}^z/2\right)\sigma^z_m}e^{i\left(\theta_{m,t}^y/2\right)\sigma^y_m}\right)$.
It is interesting to note that $U(\theta_{m,t}^{(p)})$ is only confined to invoking $\sigma^y$ and $\sigma^z$, of course, there are other ways to construct the rotations. The rotation angles $\theta_{m,t}$ are sampled from a classical distribution (such as uniform distribution, multivariate Gaussian distribution and multivariate Laplasian distribution), and the entangler $U_{ent}$ comes from the graph model $G(V,E),s.t. |V|=n$, which is uniquely determined by the structure of the RBM:
\begin{align}
U_{ent}=\prod\limits_{(i,j)\in E}|i\rangle\langle i|\otimes\sigma_j^z.
\end{align}
The implementing of the parallel hardware efficient ansatz  $W(\bm{\theta})=\sum_{p=1}^{N_h}W(\bm{\theta}^{(p)})$ depends on the Hadamard gate $H$ and controlled unitary $W=\sum_{p=1}^{N_h}|p\rangle\langle p|\otimes W(\bm{\theta}^{(p)})$, i.e.,
\begin{align}
HWH|0\rangle^{logN_h}|0\rangle^{n}=\frac{1}{N_h}|0\rangle^{\log N_h}W(\bm{\theta})|0\rangle^n+|\Psi^{\perp}\rangle,
\end{align}
where $\left(|0^{N_h}\rangle\langle0^{N_h}|\right)|\Psi^{\perp}\rangle=0$. Therefore the priori probability distribution $\rho(\mathbf{w})$ can be computed by measuring the above state on the basis $|0^n\rangle\langle0^n|$, and the oblivious amplitude amplification technique \cite{Childs2012Hamiltonian} can realize the
polynomial acceleration in this measuring step.\\
\textbf{Density matrix of likelyhood distribution.} In the feature space, the posterior distribution of RBM can be further simplified as
\begin{eqnarray}
p(\mathbf{w}|\mathcal{D})=p(\mathbf{w})\prod\limits_{l=1}^N\prod\limits_{j=1}^{N_h}\exp(1+\Phi(\mathbf{x}^{(l)})^TW_{.,j}),
\end{eqnarray}
where $W_{.,j}$ denotes the $j$-th column vector of $W$. Noting that $\Phi(\mathbf{x}^{(l)})^TW_{.,j}$ is actually an overlap, or the inner product, between the vectors $\Phi(\mathbf{x}^{(l)})$ and $W_{.,j}$ \cite{HX2014BayesianRBM}.
Given the weight state $|\mathbf{w}\rangle$ and feature state $|\Phi(\mathbf{x})\rangle_{tr}$, we attempt to estimate there overlap $\|\langle\Phi(\mathbf{x})|\mathbf{w}\rangle\|$ using quantum computer. The implementing of the following quantum algorithm depends on two two quantum techniques provided in Supplementary materials. The first technique \cite{Zhou2017Quantum}, the combination of amplitude estimation \cite{Brassard2012Quantum} and swap test, promises our proposed algorithm enabling calculating that overlap and immediately encoding it into the register, and the second one interprets the procedure of linear combination of unitary.
Drawing supportive from these two techniques, we can implement the following quantum algorithm that computes likelyhood distribution in quantum settings.\\
\textbf{Subroutine (QCL): Quantum algorithm for computing likelyhood density matrix ${\bm \rho}\left(\Phi(\mathcal{D})|\mathbf{w}\right)$}\\
\textbf{Input:} A quantum restricted boltzmann machine composed of $N_v$ visible nodes and $N_h$ hidden nodes; $n$ qubits initialized to $|0\rangle^{\otimes n}$ to encode $N_v$ visible nodes; $n$ qubits initialized to $|0\rangle^{\otimes n}$ to encode $N_h$ hidden nodes; an auxiliary qubit; and training data set $\mathcal{D}=\{\mathbf{x}^{(l)}=(\mathbf{x}^{(l)}_{data}, \mathbf{x}^{(l)}_{label} )\}_{l=1}^{N}$.\\
\textbf{Output:}Likely-hood distribution ${\rm \rho}\left(\Phi(\mathcal{D})|\mathbf{w}\right)$
\begin{enumerate}
  \item For any training data $\mathbf{x}^{(l)}\in \mathcal{D}$, we first construct quantum state $|\Phi(\mathbf{x}^{(l)})\rangle_{tr}$ in the first system, and then apply Hardmard gate $H$ to the auxiliary qubit (the second system) resulting in the state of the whole system
      \begin{align}
      \frac{1}{\sqrt{2}}|\Phi(\mathbf{x}^{(l)})\rangle_{tr}\left(|0\rangle+|1\rangle\right)|0\rangle^{\otimes n}.
      \end{align}
     The operation $|0\rangle\langle0|\otimes \mathcal{U}_{data}\mathcal{U}_{label}+|1\rangle\langle1|\otimes I$ is performed onto the the third system controlled by the second register, then we obtain
       \begin{align}
      \frac{1}{\sqrt{2}}|\Phi(\mathbf{x}^{(l)})\rangle_{tr}\left(|0\rangle|\Phi(\mathbf{x}^{(l)})\rangle_{tr}+|1\rangle|0\rangle^{\otimes n}\right).
      \end{align}
  \item According to the linear combination of unitary (LCU) technique, we have
  \begin{align}
  \frac{1}{\sqrt{2N}}\sum\limits_{l=1}^N|\Phi(\mathbf{x}^{(l)})\rangle_{tr}\left(|0\rangle|\Phi(\mathbf{x}^{(l)})\rangle_{tr}+|1\rangle|0\rangle^{\otimes n}\right).
  \end{align}
  \item Considering the QRBM model with $N_h$ hidden nodes, for the first hidden node $h_1$ and corresponding weight states $ |\mathbf{w}^{(1)}\rangle=W(\bm{\theta}^{(1)})|0\rangle^{\otimes n}$. Construct the controlled operator $|0\rangle\langle0|\otimes I+|1\rangle\langle1|\otimes W(\bm{\theta}^{(1)})$ and apply it on the second and the third system, where the parameter $\bm{\theta}^{(1)}$ is determined by the previous phase. The resulting state becomes
  \begin{align}
  \frac{1}{\sqrt{2N}}\sum\limits_{l=1}^N|\Phi(\mathbf{x}^{(l)})\rangle_{tr}\left(|0\rangle|\Phi(\mathbf{x}^{(l)})\rangle_{tr}+|1\rangle|\mathbf{w}^{(1)}\rangle \right).
  \end{align}
  \item Perform Hadamard gate $H$ on the second system, then the system turns to the following state by invoking the analogue of swap test technique (see Supplementary material), s.t.
  \begin{align}
  \frac{1}{\sqrt{N}}\sum\limits_{l=1}^{N}|\Phi(\mathbf{x}^{(l)})\rangle_{tr}|\phi(l,1)\rangle|K(\mathbf{x}^{(l)},\mathbf{w}^{(1)})\rangle,
  \end{align}
  where $|\phi(l,1)\rangle\doteq\frac{1}{\sqrt{2}}\left(|0\rangle|\Phi(\mathbf{x}^{(l)})\rangle_{tr}+|1\rangle|\mathbf{w}^{(1)}\rangle \right)$.
  \item Perform exponential operator $e^{-x}$ onto the third system, then add ancillary qubit $|0\rangle$ and utilize controlled rotation procedure, s.t. $\sum_{\theta\in\{0,1\}^t}|\theta\rangle\langle\theta|\otimes\exp(i\theta\sigma_y)$, so the state becomes
  \begin{align}
  \frac{\sum\limits_{l=1}^{N}e^{-K(\mathbf{x}^{(l)},\mathbf{w}^{(1)})}|\Phi(\mathbf{x}^{(l)})\rangle_{tr}|\phi(l,1)\rangle|K(\mathbf{x}^{(l)},\mathbf{w}^{(1)})\rangle}{\sqrt{\sum\limits_{l=1}^Ne^{-2K(\mathbf{x}^{(l)},\mathbf{w^{(1)}})}}}.
  \end{align}
  after measuring ancillary qubit $|0\rangle$ with $p(0)=\sqrt{\sum_{l=1}^N\exp(-2K(\mathbf{x}^{(l)},\mathbf{w}^{(1)}))/N}$. The parameter $t$ is the number of qubits to describe the overlap $K(\mathbf{x}^{(l)},\mathbf{w}^{(1)})$, and the controlled rotation therefore takes complexity of $\mathcal{O}(t)$.
  \item Uncomputing the step 3 and step 4, we have the state
  \begin{align}
  \frac{\sum\limits_{l=1}^{N}e^{-K(\mathbf{x}^{(l)},\mathbf{w^{(1)}})}|\Phi(\mathbf{x}^{(l)})\rangle_{tr}\left(|0\rangle|\Phi(\mathbf{x}^{(l)})\rangle_{tr}+|1\rangle|0\rangle^{\otimes n}\right)}{\sqrt{2\sum\limits_{l=1}^Ne^{-2K(\mathbf{x}^{(l)},\mathbf{w^{(1)}})}}}.
  \end{align}
  Then repeat the steps 3-5 adding the hidden nodes $h_2,...,h_{N_h}$ respectively, we finally obtain the likely-hood state of the training data $\mathcal{D}$ under the quantum restricted boltzmann machine model.
  \begin{align}
  \label{eq:likely-hood}
  |P\left(\Phi(\mathcal{D})|\mathbf{w}\right)\rangle=\frac{\sum\limits_{l=1}^{N}P(\mathbf{x}^{(l)}|\mathbf{w})|\Phi(\mathbf{x}^{(l)})\rangle}{\sqrt{\sum_{l=1}^{N}P(\mathbf{x}^{(l)}|\mathbf{w})}},
  \end{align}
  in which the probability parameter $P(\mathbf{x}^{(l)}|\mathbf{w})=\prod_{k=1}^{N}\exp\{-K(\mathbf{x}^{(l)},\mathbf{w}^{(k)})\}$ is the likely-hood probability of the data $\mathbf{x}^{(l)}$. We therefore obtain the likely-hood distribution density operator $\rho\left(\Phi(\mathcal{D})|\mathbf{w}\right)=|P\left(\Phi(\mathcal{D})|\mathbf{w}\right)\rangle\langle P\left(\Phi(\mathcal{D})|\mathbf{w}\right)|$.
\end{enumerate}
To save the sources of qubits, actually, one can also measure the third system and save each $K(\mathbf{x}^{(l)},\mathbf{w}^{(k)})$ on the classical memory without utilizing amplitude estimation technique, and finally designs the unitary operator
\begin{align}
\sum_{l=1}^NP(\mathbf{x}^{(l)}|\mathbf{w})\mathcal{U}_{\mathbf{x}_{data}}
\end{align}
and perform it on the initial state $|0\rangle^{\otimes n}$ to achieve the state of Eq.\eqref{eq:likely-hood} when implementing the experiments. Thus the procedure does not depend on the auxiliary qubit and amplitude estimation algorithm necessarily, furthermore this schema does not submerge the quantum supremacy taken by the feature state.
On the other hand, we should also point that the number of hidden nodes $N_h=\mathcal{O}(1)$ is a constant which does not depend on the scale of visible data set $\mathcal{D}$. In fact, our experiment only sets 1 hidden nodes but manifests a descent performance.\\
\textbf{Theorem 1:} Suppose the ansatz $\rho(\mathbf{w})$ can be generated in the way of $\rho(\mathbf{w})=W(\bm{\theta})\left(|0\rangle\langle0|\right)^{\otimes n}W^{\dag}(\bm{\theta})$, and there are enough layers in the quantum circuit $W(\bm{\theta})$, s.t. $W(\bm{\theta})|0\rangle^{\otimes n}$ can generate an arbitrary state, then the physical quantity ${\rm Tr}\left(\rho(\Phi(\mathcal{D})|\mathbf{w}(\bm{\theta}))\rho(\mathbf{w}(\bm{\theta}))\right)$ can be recognized as a measure to represent posterior distribution in quantum settings.\\
\textbf{Proof:} If we expand the physical quantity under the computational basis, then we have
\begin{align}
{\rm Tr}\left(\rho(\Phi(\mathcal{D})|\mathbf{w}(\bm{\theta}))\rho(\mathbf{w}(\bm{\theta}))\right)={\rm Tr}\left(\sum\limits_{i,j}p_{i,j}(\Phi(\mathcal{D})|\mathbf{w})p_{i,j}(\mathbf{w})|i\rangle\langle j|\right)=\sum\limits_{i}p_{i,i}(\Phi(\mathcal{D})|\mathbf{w})p_{i,i}(\mathbf{w}).
\end{align}
If we traverse all the possible $\mathbf{w}$, this physical quantity can be simplified as:
\begin{align}
&\frac{1}{2^n}\int_{\mathbf{w}}{\rm Tr}\left(\rho(\Phi(\mathcal{D})|\mathbf{w})\rho(\mathbf{w})\right){\rm d}\mathbf{w}=\frac{1}{2^n}\int_{\mathbf{w}}\sum\limits_{i}p_{i,i}(\Phi(\mathcal{D})|\mathbf{w})p_{i,i}(\mathbf{w}){\rm d}\mathbf{w}\\
=&\frac{1}{2^n}\sum\limits_{i}\int_{\mathbf{w}}p_{i,i}(\Phi(\mathcal{D})|\mathbf{w})p_{i,i}(\mathbf{w}){\rm d}\mathbf{w}=\frac{1}{2^n}\sum\limits_{i}\int_{\mathbf{w}}p_{i,i}(\mathbf{w}|\Phi(\mathcal{D})){\rm d}\mathbf{w}\doteq1
\end{align}
Therefore, the physical quantity, s.t. the overlap between likelyhood and priori density matrixes, can be recognized as a measure to represent posterior distribution in quantum settings.\\
\textbf{Theorem 2:} If the ansatz $\rho(\mathbf{w})$ can be prepared by $\rho(\mathbf{w})=W(\bm{\theta})\left(|0\rangle\langle0|\right)^{\otimes n}W^{\dag}(\bm{\theta})$, then modifying the rotation parameter $\bm{\theta}$ can actually traverse all the possible weight parameter $|\mathbf{w}\rangle$ with an affordable error. \\
\textbf{Proof:} Suppose we perform a sequence $V_1,...,V_m$ of gates intended to approximate some other sequence of gates: $U_1,...,U_m$. Then it turns out that the error caused by the entire sequence of imperfect gates is at most the sum of the errors in the individual gates:
\begin{align}
E\left(U_mU_{m-1}U_1,V_mV_{m-1}...V_1\right)\leq \sum\limits_{j=1}^mE(U_j,V_j).
\end{align}
Now we consider the local single qubit rotation $U^{(t)}_{loc}(\theta_t)=\bigotimes_{m=1}^n\left(e^{i\theta_{m,t}^zZ_m}e^{i\theta_{m,t}^yY_m}\right)$, then the difference between $U^{(t)}_{loc}(\theta_t)$ and $U^{(t)}_{loc}(\theta_t+\Delta_{\mathbf{w}})$ can be measured as
\begin{align}
\|U^{(t)}_{loc}(\theta_t)-&U^{(t)}_{loc}(\theta_t+\Delta_{\mathbf{w}})\|
=\|\bigotimes\limits_{m=1}^n\left((e^{i\theta_{m,t}^zZ_m}-e^{i(\theta_{m,t}^z+\Delta_{\mathbf{w}})Z_m})|0\rangle\langle0|+(e^{-i\theta_{m,t}^zZ_m}-e^{-i(\theta_{m,t}^z+\Delta_{\mathbf{w}})Z_m})|1\rangle\langle1|\right)\\
&\left((e^{i\theta_{m,t}^yY_m}-e^{i(\theta_{m,t}^y+\Delta_{\mathbf{w}})Y_m})|+\rangle\langle+|+(e^{-i\theta_{m,t}^yY_m}-e^{-i(\theta_{m,t}^y+\Delta_{\mathbf{w}})Y_m})|-\rangle\langle-|\right)\|
\end{align}
\begin{align}
\leq\|\bigotimes\limits_{m=1}^n\left(i\Delta_{\mathbf{w}}U^{(t)}_{loc}(\theta_t)\right)\|=\left(\Delta_{\mathbf{w}}\right)^n\|U^{(t)}_{loc}(\theta_t)\|
\end{align}
Combining the theorem, therefore the difference between  $|\mathbf{w}^{(k)}\rangle$ and $|\mathbf{w}^{(k+1)}\rangle$ can be measured as
\begin{align}
E(W(\theta+\Delta_{\mathbf{w}}),W(\theta))\leq\sum\limits_{l=1}^L\left(\Delta_{\mathbf{w}}\right)^n\|U^{(l)}_{loc}(\theta_l)\|=\mathcal{O}(L\left(\Delta_{\mathbf{w}}\right)^n).
\end{align}
Thus the microelement $d\mathbf{w}=d(\bm{\theta}^n)$, which implies the fact that modifying the rotation parameter $\bm{\theta}$ can actually traverse all the possible weight states with an exponential small error. \\


\section*{Acknowledgments}
This work is supported by NSFC (Grant Nos. 61672110, 61671082, 61976024, 61972048), and the Fundamental Research Funds for the Central Universities (Grant No.2019XD-A01). We thank Bailing Zhang, Shijie Pan and Binbin Cai on discussing the writing thoughts, and we also acknowledge using of the HiQ for this work.
\section*{Author contributions}

Y.W. contributed to the initiation of the idea. All authors wrote and reviewed the manuscript.
%
\end{document}